\documentclass[final,5p,times,twocolumn,sort&compress]{elsarticle}
\usepackage{amssymb}
\usepackage{amsmath}
\usepackage{epstopdf}
\usepackage{graphicx}
\usepackage{lineno}
\journal{Nucl. Instr. Meth. A}
 \def\ds{\displaystyle}
 \allowdisplaybreaks
\begin{document}
 \begin{frontmatter}
 \title{AWAKE-related benchmarking tests for simulation codes}
 \author[binp,nsu]{K.V.~Lotov}
 \address[binp]{Budker Institute of Nuclear Physics SB RAS, 630090, Novosibirsk, Russia}
 \address[nsu]{Novosibirsk State University, 630090, Novosibirsk, Russia}

 \date{\today}
 \begin{abstract}
Two tests are described that were developed for benchmarking and comparison of numerical codes in the context of AWAKE experiment.
  \end{abstract}
 \begin{keyword}
Plasma wakefield acceleration \sep
Proton driver \sep
Simulations
 \end{keyword}
 \end{frontmatter}


\section{Introduction}

AWAKE project at CERN is currently one of flagship experiments on particle beam-driven plasma wakefield acceleration and the only experiment with proton drivers \cite{NIMA-829-3,NIMA-829-76,PPCF56-084013}. It aims at better understanding the physics of the acceleration process, demonstrating high-gradient acceleration with a proton bunch, and developing necessary technologies for the long-term perspectives of proton-driven plasma wakefield acceleration.

Studies of AWAKE physics present a real challenge for simulation codes. This is because parameters of the experiment fall far beyond the area for which most codes were originally developed and tuned. Proton beams are very long, up to several hundreds plasma wavelengths. The excited wakefield grows nearly exponentially in space and time and depends on a small-amplitude seed perturbation. At the same time, the design of the experiment relies mainly on simulation results rather than on earlier experiments, as this is the first experiment with seeded self-modulation of a proton bunch. Therefore, it is important to provide a thorough validation of available codes at test problems that contain the main physical effects involved.

\begin{figure*}[tb]\centering
 \includegraphics[width=288bp]{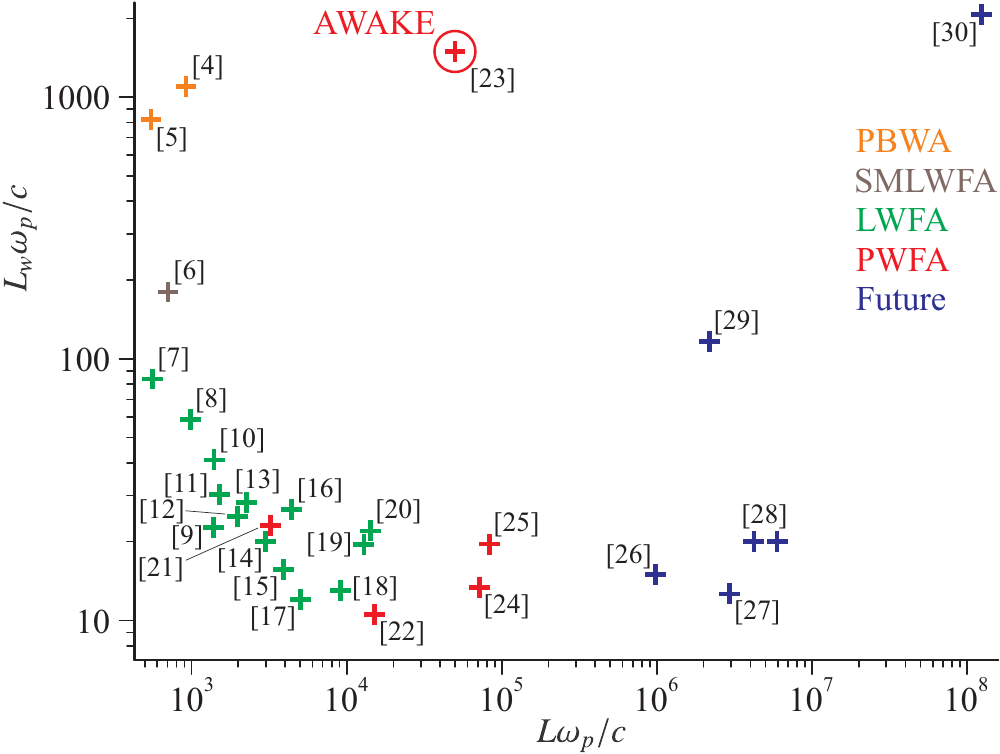}
\caption{Simulation scales of key laser- and beam-driven plasma wakefield experiments and future-concept theoretical studies. The data are taken from papers referenced near the points. Colors are used to distinguish experiments with different drivers and theoretical studies of future projects.}\label{fig1-map}
\end{figure*}

Figure~\ref{fig1-map} illustrates how far AWAKE parameters are from the well-studied area. Here we show the required length of the simulation window $L_w$ and the beam propagation distance $L$, both measured in units of $c/\omega_p$, where $c$ is the speed of light, and $\omega_p = \sqrt{4 \pi n e^2/m}$ is the plasma frequency determined by the plasma density $n$, elementary charge $e$, and electron mass $m$. AWAKE makes a two orders of magnitude jump in beam length up from the area where the codes were recently benchmarked against experiments. Future applications of proton drivers make additional four orders of magnitude jump in propagation distance.

In this paper, we describe two tests for benchmarking and comparison of numerical codes that were used to simulate beam-plasma interaction in the AWAKE experiment. Most of expected test results are illustrated with high-resolution runs of quasi-static 2d-axisymmetric kinetic code LCODE \cite{PRST-AB6-061301,NIMA-829-350}. Test results of several other codes can be found in Refs.~\cite{NIMA-829-3,IPAC13-1238}.

\section{Test 1: Long term behavior of a small-amplitude plasma wave}

Here we follow the long term evolution of a small amplitude plasma wave generated by a short proton bunch. This test shows quality of plasma and field solvers and also helps to determine the required run resolution and number of plasma macro-particles. A small perturbation that seeds beam self-modulation in AWAKE must maintain a constant amplitude over hundreds of wave periods. The wave period must be simulated correctly to a fraction of percent, as the wave phase at the witness location would be wrong otherwise \cite{PoP21-083107}. Consequently, we focus at conservation of the wave amplitude and at the wave period.

\begin{figure}[tb]\centering
\includegraphics[width=241bp]{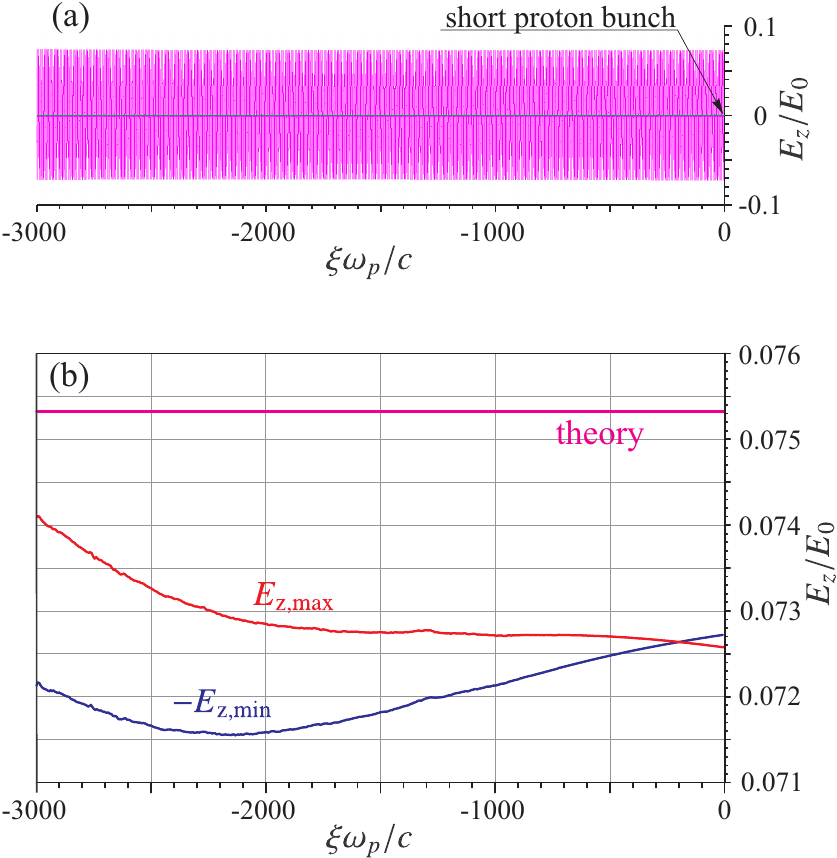}
\caption{On-axis electric field $E_z$ in test 1: (a) general view, (b) zoomed-in amplitude variations measured with positive ($E_\text{z,max}$) and negative ($-E_\text{z,min}$) local extrema.}\label{fig2-t1ampl}
\end{figure}

\begin{figure}[tb]\centering
\includegraphics[width=241bp]{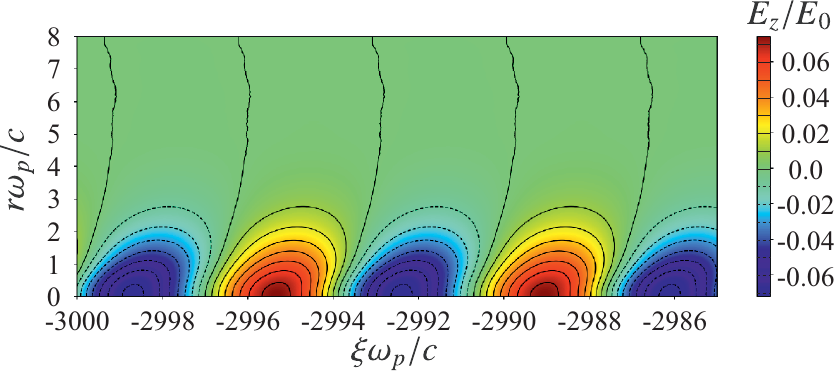}
\caption{Wave shape at the end of the simulation area.}\label{fig3-t1wave}
\end{figure}

The proton beam density is
\begin{equation}\label{e1}
 n_b = \left\{ \begin{array}{ll}
 \ds 0.5 \, n_{b0} \, e^{-r^2/2 \sigma_r^2} \left[  1 + \cos \left(  \sqrt{\frac{\pi}{2}} \frac{\xi}{\sigma_z}  \right)  \right], & |\xi| < \sigma_z \sqrt{2\pi}, \\
 0, & \text{otherwise},
 \end{array}\right.
\end{equation}
where we use cylindrical coordinates $(r, \phi, z)$ and the co-moving coordinate $\xi = z-ct$; the $z$-axis is the direction of beam propagation. The beam parameters are
\begin{equation}\label{e2}
  \sigma_r=c/\omega_p, \quad \sigma_z=c/\omega_p, \quad  n_{b0}=0.1n.
\end{equation}
In this test, the proton beam is assumed unchangeable in the co-moving frame, and plasma ions are immobile. Boundaries of the simulation window must be far enough to approximate bunch propagation in an unbounded uniform plasma of the density $n$. We follow the excited wakefield at some fixed point up to the time $3000\,\omega_p^{-1}$ after the driver passage [Fig.\,\ref{fig2-t1ampl}(a)] and measure the average wave period and value of the wakefield amplitude. We also compare two-dimensional maps of the field $E_z$ on $(r,\xi)$ plane at last two periods of the simulated wave (Fig.\,\ref{fig3-t1wave}).


To produce Figures~\ref{fig2-t1ampl} and \ref{fig3-t1wave} with the best quality, we use square grid of the size $dr = d\xi = 0.0025c/\omega_p$, $40\,000$ macro-particles for plasma electrons, simulation area of the radius $R = 16 c/\omega_p$, and analytical definition of the  driver current (as opposed to forming an ensemble of individual macro-particles). The run takes about 20 CPU hours. Lower resolution, 32 times faster runs with $dr = d\xi = 0.01c/\omega_p$, $R = 8 c/\omega_p$ can also produce acceptable results \cite{NIMA-829-3}.

According to the linear wakefield theory \cite{PAcc20-171}, the longitudinal electric field on the axis must have the oscillation amplitude $E_\text{z,max} = 0.07532376924\,E_0$, where $E_0 = mc\omega_p/e$. This amplitude is order of magnitude larger than that of the seed perturbation in the baseline AWAKE scenario \cite{PoP21-123116}; it was chosen so high to ease its measuring in the presence of numerical noise. However, nonlinear effects are visible at this driver density. The simulated amplitude is initially 3.5\% lower than the theory value [Fig.\,\ref{fig2-t1ampl}(b)]. This is a first-order effect in drive beam density, and for a negatively charged bunch, the amplitude is approximately 3\% higher (not shown in the figure). Small variations of the amplitude along the simulation window  [Fig.\,\ref{fig2-t1ampl}(b)] have no theoretical explanation yet, so we cannot conclude whether they are physical or numerical effect.

The wave period is also modified by nonlinear effects \cite{PoP20-083119}. For the selected driver radius, it is
\begin{equation}\label{e3}
    \tau \approx 2 \pi \omega_p^{-1} \left(1 + 0.10 (E_\text{z,max}/E_0)^2\right),
\end{equation}
so the period is longer than $2 \pi \omega_p^{-1}$ by 0.053\%. Our simulations agree with this value.

\begin{figure*}[tb]\centering
\includegraphics[width=483bp]{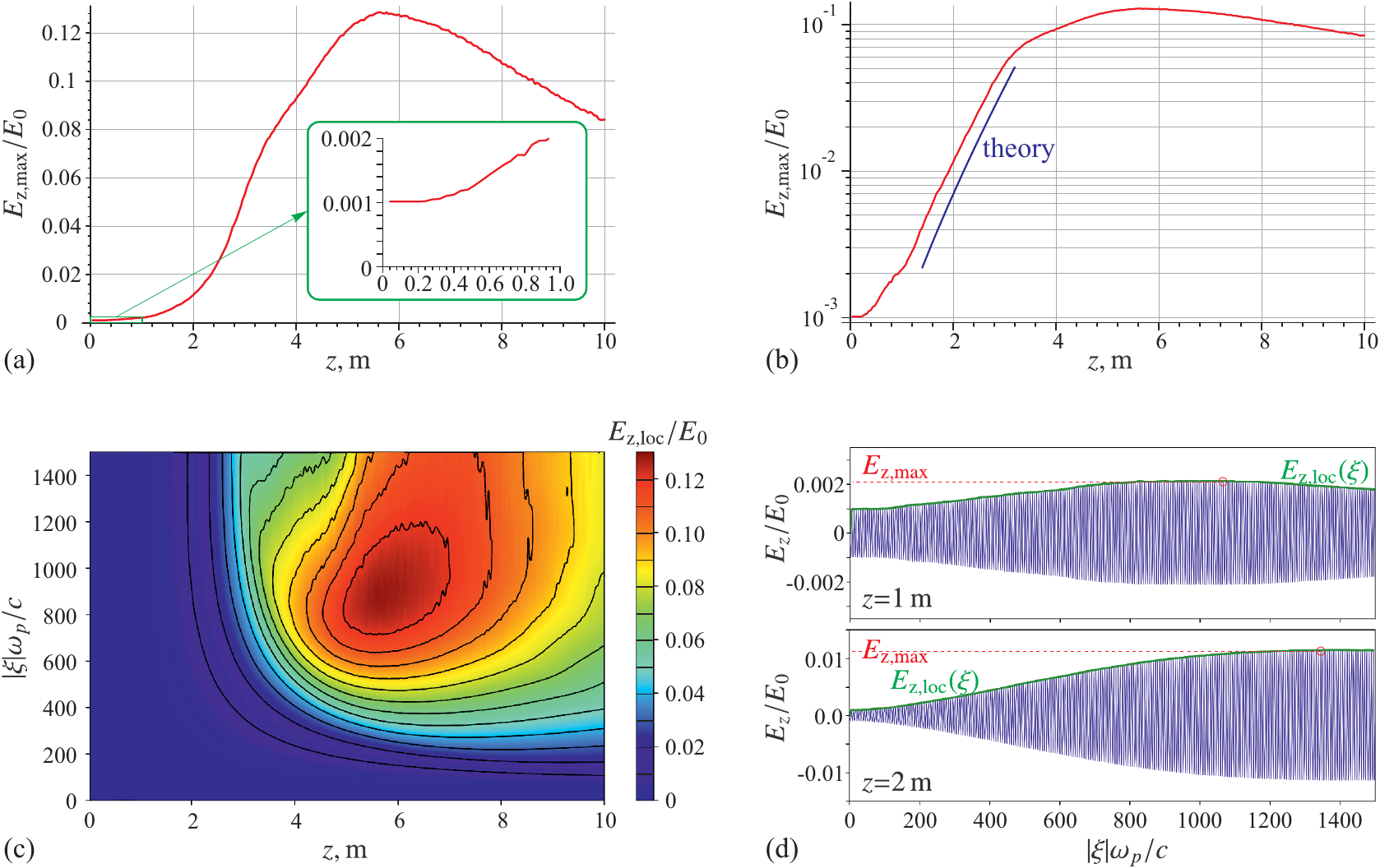}
\caption{The maximum wave amplitude $E_\text{z,max}$ versus propagation distance $z$ in (a) normal and (b) semi-logarithmic scales. The blue line in (b) shows the theoretically predicted growth and is shifted vertically for better visibility. (c) Amplitudes of local field maxima $E_\text{z,loc} (z, \xi)$. (d) Time evolution of the wakefield $E_z(\xi)$ at two selected cross-sections.}\label{fig4-test2}
\end{figure*}

If the drive bunch is composed of individual macro-particles, they produce a wakefield noise \cite{PRST-AB16-041301} that can be additionally amplified by periodic boundary conditions \cite{PoP24-103129}. The wakefield noise does not look like a usual plasma noise, but manifest itself as some random addition to the wakefield amplitude. According to formula (10) of Ref.~\cite{PRST-AB16-041301}, the driver \eqref{e1}--\eqref{e2} composed of $N$ randomly distributed macro-particles produces a noise wakefield of the amplitude
\begin{equation}\label{e4}
    E_{az} \sim 0.1 E_0 / \sqrt{N}.
\end{equation}
It is smaller than 1\% of the regular wakefield for $N > 3 \times 10^4$.

\section{Test 2: Seeded self-modulation}

In this test, we simulate development of beam self-modulation seeded by a hard leading edge of the driver. The test mimics one of early AWAKE scenarios named SPS-LHC case \cite{PoP18-103101}. There is no detailed theory of the process, so the result is evaluated by agreement with high resolution runs of several acknowledged codes \cite{NIMA-829-3}. Here we use the grid size $dr = d\xi = 0.005c/\omega_p$, 10 particles per cell ($16\,000$ macro-particles in total), simulation area of the radius $R = 8 c/\omega_p$, time step for the beam $dz/c = 100 \omega_p^{-1}$, and $4.5 \times 10^7$ macro-particles for the beam. This LCODE run  takes about 750 CPU hours.

\begin{table}[tb]
 \begin{center}
 \caption{ Test 2 parameters. Numbers in bold are exact values, others can be refined. }\label{t1}
 \begin{tabular}{lll}\hline
  Parameter and notation & Value \\ \hline
  Plasma density, $n$ & $7.06 \times 10^{14}\,\text{cm}^{-3}$ \\
  Plasma skin depth, $c/\omega_p$, & \textbf{0.02}\,cm \\
  Wavebreaking field, $E_0$, & 2.54\,GV/m \\
  Plasma length, $L$, & 10\,m \\
  Uncut beam population, $N_b$ & $\mathbf{1.15\times 10^{11}}$ \\
  Beam length, $\sigma_z$ & \textbf{12}\,cm \\
  Beam radius, $\sigma_r$ & \textbf{0.02}\,cm & \\
  Maximum beam density, $n_{b0}$ & $1.52\times 10^{12}\,\text{cm}^{-3}$ \\
  Beam energy, $W_b$ & \textbf{450}\,GeV \\
  Beam energy spread, $\delta W_b$ & \textbf{135}\,MeV \\
  Beam emittance, $\epsilon$ & $\textbf{8}\,\mu$m\,mrad \\
  Beam angular spread, $\delta \alpha = \epsilon/\sigma_r$, & $\mathbf{4 \times 10^{-5}}$ \\
  \hline
 \end{tabular}
 \end{center}
\end{table}

At the entrance to the plasma (at $z=0$) the beam density is
\begin{equation}\label{e5}
 n_b = \left\{ \begin{array}{ll}
 \ds 0.5 \, n_{b0} \, e^{-r^2/2 \sigma_r^2} \left[  1 + \cos \left(  \sqrt{\frac{\pi}{2}} \frac{\xi}{\sigma_z}  \right)  \right], & -\sigma_z \sqrt{2\pi} < \xi < 0, \\
 0, & \text{otherwise}.
 \end{array}\right.
\end{equation}
Here
\begin{equation}\label{e6}
 n_{b0} = \frac{N_b}{(2\pi)^{3/2} \sigma_r^2 \sigma_z}.
\end{equation}
Quantitative parameters of the system are given in Table~\ref{t1}. The plasma is uniform, plasma ions are immobile. Beam protons have the following momentum distribution:
\begin{multline}\label{e7}
  f_b(p_x,p_y,p_z) =  \frac{c^3}{(2\pi)^{3/2} \, \delta \alpha^2 \, W_b^2 \, \delta W_b} \\
  \times \exp \left( -\frac{(p_x^2+p_y^2) c^2}{2 \, \delta \alpha^2 W_b^2} - \frac{(p_z c-W_b)^2}{2 \, \delta W_b^2} \right).
\end{multline}

We compare the maximum wakefield amplitude $E_\text{z,max}$ excited at various $z$  [Fig.\,\ref{fig4-test2}(a,b)], no matter what time after beam head passage this field is reached. Before the beam shape changes (at $z=0$), the wakefield amplitude can be calculated analytically and equals $9.94 \times 10^{-4}E_0$ [the inset in Fig.\,\ref{fig4-test2}(a)]. Amplitudes of local field maxima $E_\text{z,loc}$ as functions of propagation distance ($z$) and delay with respect to the bunch head ($|\xi|$) give more information on the structure of excited fields [Fig.\,\ref{fig4-test2}(c)], but are less convenient for comparison. Figure~\ref{fig4-test2}(d) illustrates how the quantities $E_\text{z,max}$ and $E_\text{z,loc}$ are measured.

The theory of seeded self-modulation is only available for narrow ($\sigma_r \ll c/\omega_p$) beams of a constant current \cite{PRL107-145003}. Nevertheless, the theoretically predicted growth
\begin{equation}\label{e8}
    E_\text{z,max} \propto \exp \left( \frac{3 \sqrt{3}\omega_p L^{1/3} z^{2/3}}{2^{7/3} c} \left( \frac{n_{b0} m c^2}{n W_b} \right)^{1/3} \right)
\end{equation}
closely match simulations, if we take $\sigma_z$ as the beam length $L$ [Fig.\,\ref{fig4-test2}(b)].

\section{Acknowledgements}

This work is supported by The Russian Science Foundation, grant No.~14-50-00080. Computer simulations are made at Siberian Supercomputer Center SB RAS.

\end{document}